\documentclass[conference]{IEEEtran}
\IEEEoverridecommandlockouts
\usepackage{cite}

\ifCLASSINFOpdf
  \usepackage[pdftex]{graphicx}
  \graphicspath{{../pdf/}{../jpeg/}}
  \DeclareGraphicsExtensions{.pdf,.jpeg,.png}
\else
  \usepackage[dvips]{graphicx}
  
  \graphicspath{{../eps/}}
\fi
\usepackage{subfigure}

\usepackage[utf8]{inputenc}
\usepackage[english]{babel}
\usepackage{lastpage}
 
\begin{document}
%
\title{An Energy Packet Switch for Digital Power Grids}
\author{\IEEEauthorblockN{ Roberto Rojas-Cessa,\thanks{This material is based upon work supported by the National Science Foundation under Grant No. (CNS) 1641033.} Chuan-Kuo Wong\thanks{This is a version of paper presented in \cite{eps-ithings18}.}, Zhengqi Jiang,\\ Haard Shah, Haim Grebel}
\IEEEauthorblockA{Department of Electrical and Computer Engineering\\
New Jersey Institute of Technology\\
Newark, NJ, USA 07102\\
Email: rojas@njit.edu}
\and
\IEEEauthorblockN{Ahmed Mohamed}
\IEEEauthorblockA{Department of Electrical Engineering\\
City College of City University of New York\\
New York, NY, USA 11031\\
Email: amohamed@ccny.cuny.edu
}
}

\pagenumbering{arabic}

\maketitle

\begin{abstract}

We propose the design and electrical description of an energy packet switch for forwarding and delivery of energy in digital power grids in this paper. The proposed switch may receive energy from one or multiple power sources in the form of energy packets, store them and aggregate the contained energy, and forward the accumulated energy to requesting loads connected to one or multiple output ports of the switch.  Energy packets are discrete amounts of energy that are associated in- or out-of-band with an address and other metadata. Loads receive these discrete amounts of finely-controlled energy rather than discretionary amounts after. The control and management of the proposed switch are based on a request-grant protocol. Using energy packets helps to manage the delivery of power in a reliable, robust, and function form that may enable features not yet available in the present power grid.  The switch, as any element of a digital grid, uses a data network for the transmission of these requests and grants. The energy packet switch may be the centerpiece for creating infrastructure in the realization of the digital power grid. The design of the energy packet switch is based on shared supercapacitors to shape and manage discretization of energy. We introduce the design and analysis of the electrical properties of the proposed switch and describe the procedure used in the switch to determine the amount of energy transmitted to requesting loads.
\end{abstract}


%
\IEEEpeerreviewmaketitle


\section{Introduction}

During the past decades, the North American power infrastructure has evolved into what many experts consider to be the largest and most complex system of the technological age. However, the vulnerability and potential problems of the power grid have placed the challenges of energy transmission and distribution into the limelight. Recently, the concept of a digital grid (DG) has been proposed \cite{saitoh1995new, abe2011digital, xu2012allocation}. In such a paradigm, energy is transmitted through the grid as data is transmitted through the Internet. Elements of the grid (e.g., generators, distributors, buses, and loads are interconnected through the Internet) play active roles in the estimating and configuring the path electrical energy follows from energy generators to consuming loads. Our approach to the digital grid is the controlled-delivery power grid (CDG) \cite{xu2012allocation}. This paradigm has been  proposed to perform a finer and more efficient management of energy distribution, which in turn improves the balance between generation of electrical power and the demand of it \cite{xu2012allocation, rojas2013management1, rojas2013management2, rojas2014testbed}. 

In the quest of realizing a DG, there have been many recent efforts. This grid, which is greatly inspired by the operation of the Internet \cite{saitoh1995new}, is expected to share many of the properties of a data network and with that, provide the level of service of today's power grid plus additional features needed to overcome its weaknesses. The improvements are a greater level of resiliency and direct integration of alternative energy sources. In the digital grid, energy is the analogous to what data is to the Internet. Therefore, digitization of energy is sought to complete the analogy. However, digitization of energy is a concept complex to realize. A reason for that is the existing long tradition of using the grid passively, where energy is considered a flow whose behavior adheres to Kirkhoff's laws. Nevertheless, digital energy bits may be considered as discrete amounts of energy, and a possible interpretation of it is an amount of power flowing for a period of time.

The feature of the present grid about providing discretionary access to energy demands to keep the grid perpetually energized. This uncontrolled accessibility requires generators to adapt the generation of power to the extent of consumption. Balancing the grid is such a careful and sensitive act that the incorporation of alternative energy sources with intermittent active times into the grid makes it complex and, in some cases, prohibitive.  

The capacity of energy supply under discretionary access is determined by the physical infrastructure, allowing cases of over-demand, and when it occurs, overloaded distribution feeders ought to be taken out of the grid, generating blackouts. Close monitoring of the grid's performance may be achieved by deploying (auxiliary) sensing data networks \cite{gono2007reliability, he2010toward, galli2011grid,
  liu2011consumer, budka2010geri, lu2010smartgridlab, yu2011coordinating, abe2011digital, bouhafs2012links, fan2013smart}. Concerns about ensuring working paths, yet perpetually energized, translate into additional management complexity \cite{abe2011digital, takuno2010home}. These works show how the
adoption of a controlled distribution of power may be seamlessly coupled with grid monitoring.

The DG offers an alternative for performing precise control on energy delivery. In a DG as in the CDG, users may issue requests for energy and the provider may fully or partially grant them within a period of time. Such an approach facilitates an estimation of total demand and gives the provider the ability to determine how and when to satisfy the requests. This management model also favors the adoption of a highly controlled
the supply. 

The concept of controlling the distribution of energy through micro-grids is being considered as the next generation electrical grid \cite{he2008architecture}. Approaches to verify user identification before the start of energy transmission in point-to-point communications have been proposed as part of a more advanced grid \cite{abe2011digital}. However, the ability to scale up point-to-point distribution systems is called into question. Some of these works are motivated by the consideration of multiple generators, or alternative-energy sources. In such cases, sources and appliances can be matched through dedicated lines, using direct current (DC) multiplexors \cite{takuno2010home}. However, uncontrolled delivery (and consumption) remains along with its associated challenges. Elastic loads have been proposed to balance the grid and to control energy delivery \cite{alizadeh2012packet}. However, such an approach requires scheduling of user loads be performed by the provider and not the user. 

In summary, recent efforts to define a power or energy switch have been centered on direct or alternating current controllers where paths are enabled by Internet addresses. The properties of having a permanently energized grid and discretionary loads remain in existing designs, realizing but partial digitization of the grid. These facts raise the following question: Is it possible to control the energy delivered in discrete amounts to a load on a network-controlled power grid as a more robust approach to a digital grid?

To address this question, we propose an {\it energy packet switch} that receives and supplies energy in discrete and addressable amounts. The switch receives energy by the ingress ports and issues energy by the egress ports. The combination of transmitting energy in finite and discrete amounts with associated network addresses gives place to what we call an energy packet. An energy packet is issued by the switch after the execution of a request-grant protocol, where loads request amounts of energy needed to operate before energy is actually supplied. Our energy packet design is based on limiting amounts of energy that can be delivered to a load. Rather than limiting the amount of energy as is being delivered, our design is based on limiting the total amount of energy to be delivered before transmission starts. This operation is achieved by using energy containers implemented with supercapacitors. These supercapacitors shape energy packets, enable receiving energy from multiple and diverge sources, and supply energy to one or multiple diverse loads. 

In this paper, we introduce the design of an energy packet switch (EPS) and show and discuss its properties. We also show how energy is transmitted from inputs to outputs. The proposed EPS is able to combine energy from multiple sources without affecting stability of the power loop and, therefore; the complete grid. The combination of controlled energy supply through energy packets and the use of the request-grant protocol increase reliability and reliance of the grid under challenging environments.

The remainder of this paper is organized as follows. Section \ref{sec:CDG} introduces the concept of the controlled-delivery power grid.
Section \ref{sec:testbed} introduces the proposed energy packet switch. Section \ref{sec:results} shows evaluations on experiments transferring energy from energy sources to the switch and from the switch to loads. Section \ref{sec:conclusions} presents our conclusions.

\section{Controlled-Delivery Power Grid}
\label{sec:CDG}

The main goal of a power grid with controlled delivery is to supply discrete and finite amounts of energy as a DG. The adoption of this approach may minimize the difference between energy generation and demand, facilitate the power distribution between starving and overpowered grids, and increase the stability of a power grid through distribution planning and instantaneous monitoring. Moreover, having prior knowledge of energy demand, the grid may act on overwhelming demands by providing limited energy or routing it. 

\begin{figure}[!t]
\centering
\includegraphics[width=2.8in]{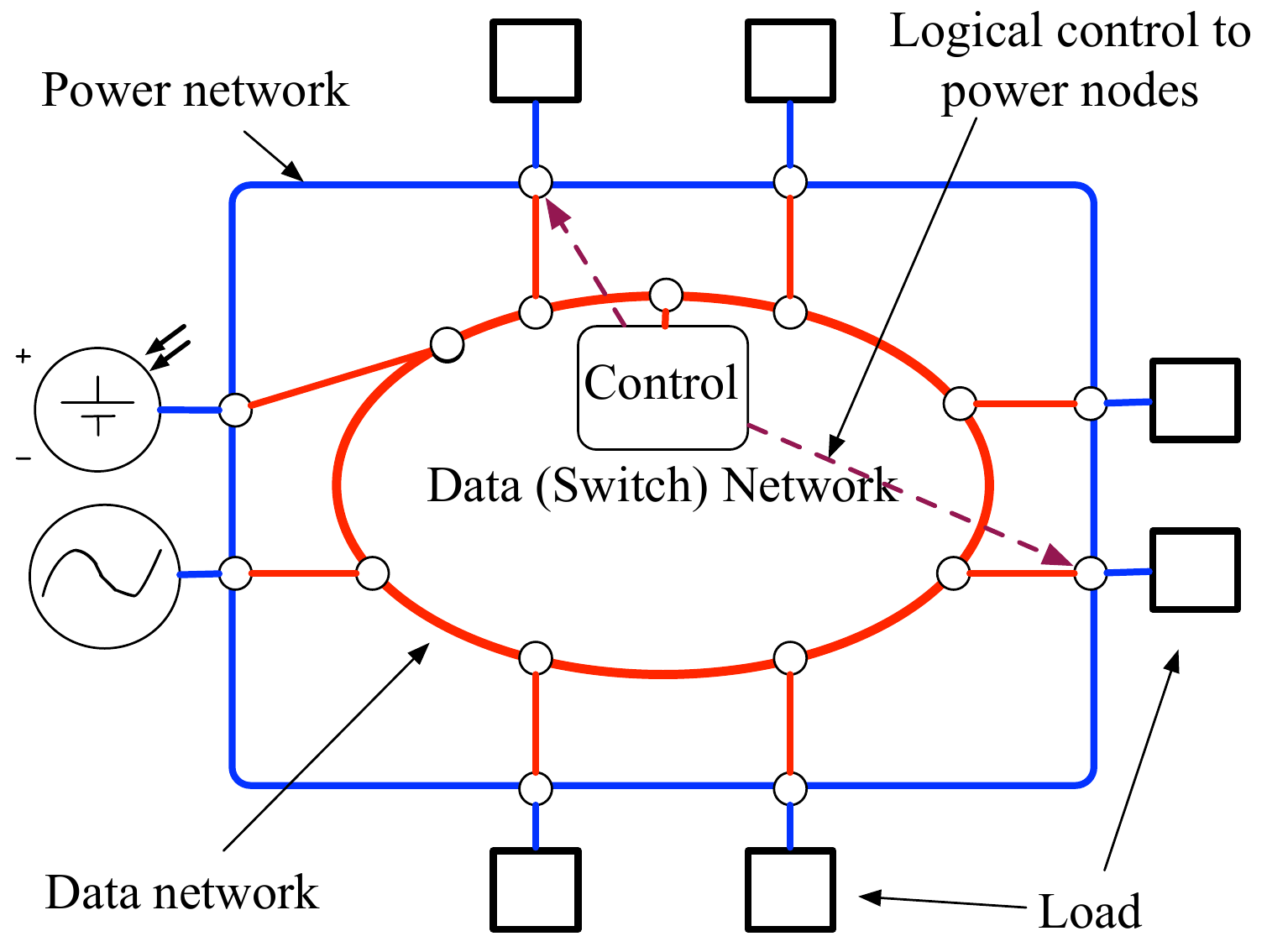}
\caption{CDG using a parallel data network for controlling power network.}
\label{fig:data-power-model}
\end{figure}

In order to overcome exposing a power grid to any number of loads, energy packets carry the destination address(es) of specific customer(s) who are the only one(s) allowed to access the transmitted energy. Internet Protocol (IP) addresses may be used in this application. Correspondingly, each load and generator has an identification number or IP address.  The assignment of addresses to users (and components of the grid) enables energy ownership to users or loads. The destination address(es) may be embedded on the electrical signal(s) or sent through a parallel data network. Figure \ref{fig:data-power-model} shows an example of the CDG using a parallel data network. Data is exchanged between loads and generators to determine demand and supply. 
 
A power access point (PAP) at the customer's premises, which is a piece of CDG equipment used by a customer as interface with the power grid and data network, enables a user access to granted power if the address carried in the signal matches that of the customer. 

In this paper, we consider that time is slotted, where the duration of a {\it time slot} is determined by the time it takes to supply the granted energy from the source to the load. The duration of a time slot determines the duration of an inter-packet gap on the parallel data network for data and power synchronization. The power network is then synchronized to the data network, while still having the overall power network operate asynchronously.

The amount of energy delivered per slot may be scaled up in two dimensions: 1) by assigning a number of time slots, back-to-back, as a train of granted requests where each time slot carries a fixed amount of energy, or 2) by setting the amount of energy transmitted within a time slot through adjustment of current and voltage.  In this paper, we fix the voltage and adjust current in a time slot. 

Energy delivery in the CDG follows a request-grant protocol performed between loads and a generator/supply. After being requested by the customer(s), energy is then supplied in compliance with a policy, which may be shaped by the physical, economical, and management limits of the feeder, to one or a large number of users. The energy supplier embeds addresses and the amount of energy granted per user in each grant. 
In the communication performed by the demanding loads and supply, the energy source:
 a) finds the requested energy demands as issued by loads and assigns energy coming from energy
sources to supply those requests, b) finds routing information about where to forward the energy, and c) issues grants and dispatches the granted energy. The EPS plays the role of a load when receiving energy from sources and the role of source when transferring energy to loads.

In this paper, we consider that a user may be able to receive energy packets at the same rate at which the EPS issues or receives energy. In other words, a user may be equipped with similar energy storage units as a switch such that the user may be able to receive e integrate energy sent intermittently by the switch. The user's PAP communicates with the EPS to request energy and receives grants from it.  

\section{Energy Packet Switch (EPS)}

\begin{figure}[htb]
\centering
\includegraphics[width=3.5in]{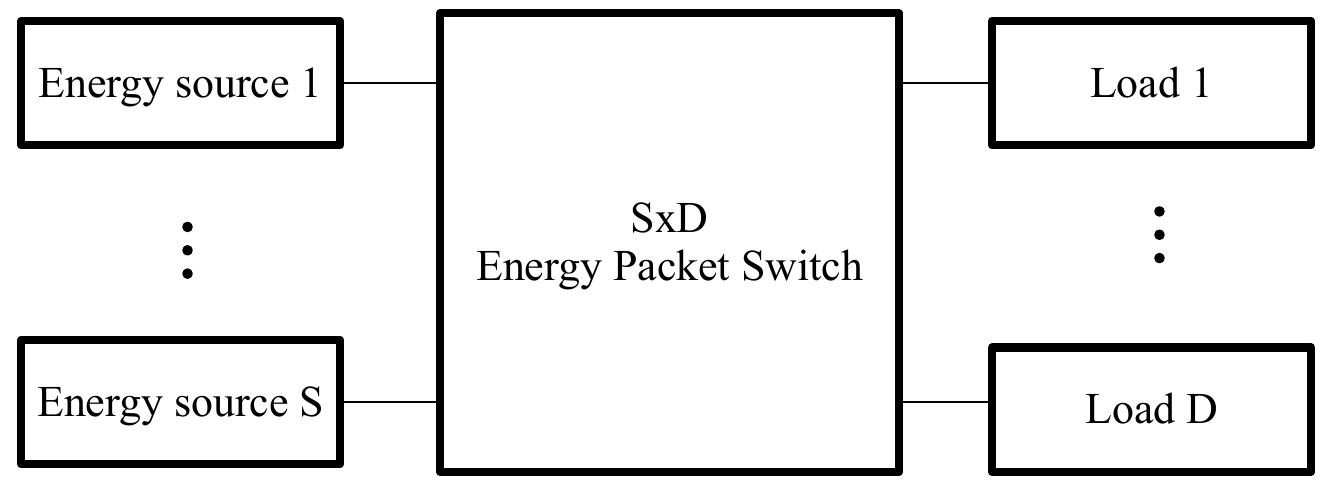}
\caption{Example of application of the energy packet switch.}
\label{fig:energy-packet-switch-application}
\end{figure}

\begin{figure}[htb]
\centering
\includegraphics[width=3.5in]{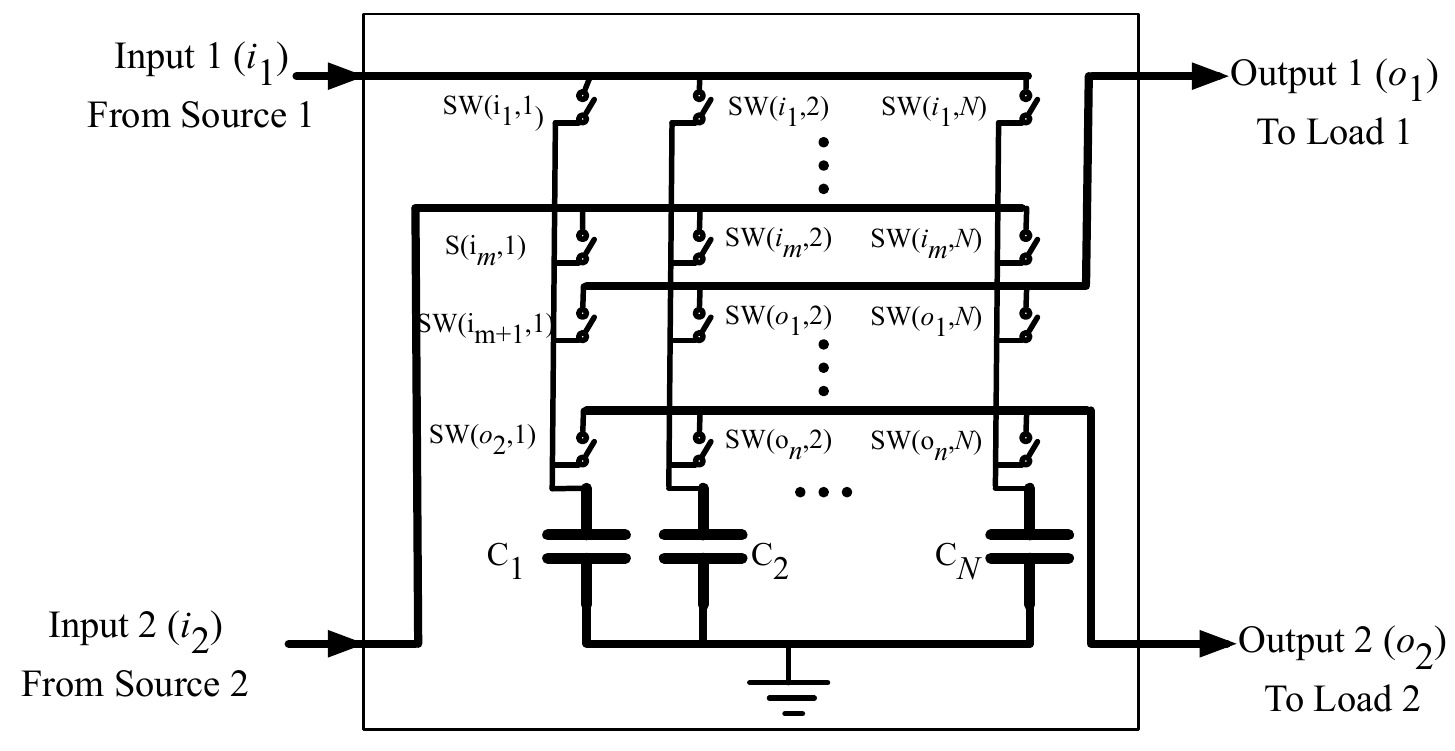}
\caption{Schematic of the energy packet switch using super-capacitors as energy buffers.}
\label{fig:energy-packet-switch}
\end{figure}

The EPS is a network-controlled switch that has $S$ inputs and $D$ outputs. Inputs serve to connect energy sources (or another EPS playing that role) to the EPS  and the outputs serve to supply the energy to requesting loads. Figure \ref{fig:energy-packet-switch-application} shows an example application of EPS where the switch is used to integrate the energy provided by $S$ different energy sources and supply energy to up to $D$ different feeders (or busses) to where one or multiple loads are connected. 
The switch is based on multiple units of shared energy storage, where one or multiple sources connected to the inputs of the switch may supply energy to one or multiple storage units. The single or multiple storage unit stores energy for short terms and receive and supply energy to loads at fast rates as well. In this version of the EPS, a shared storage is implemented with a supercapacitor and it works with direct current (DC). Figure \ref{fig:energy-packet-switch} shows a basic schematic of a 2$\times$2 (2 inputs, 2 outputs) EPS. In the figure, the switching elements connecting an input (or output) to a capacitor allow controlled access to energy sources by the capacitors. To be able to interface the EPS with loads for a proper energy transfer, we consider that loads use a supercapacitor as interface. In this way, supercapacitors can transfer energy at a proper rate and for a controllable time. The switching elements are implemented with solid-state relays (SSRs).  On the other hand, outputs also can tap to one or multiple supercapacitors to transfer energy from them to one or multiple loads. By using requests (i.e., energy demand) and the capacity of supercapacitors, the EPS determines how much energy is obtained from an energy source per exchange cycle, or time slot. 

Inputs share the storage for energy sent to the switch. Similarly, one or more loads may receive energy from the shared storage. The energy received (and sent) by the switch is limited by the capacity of the supercapacitors, and therefore, the exchange of energy from a source or to a load is a discrete and limited amount. For example, when a source supplies energy to it, the switch  cannot take more than what is requested (or admissible by the switch's capacity). Similarly, loads are not supplied more energy than that set by the capacity of the energy stored by the switch. The property of supplying discrete amounts of energy differentiates EPS from present distribution points in the present power grid, where loads may take a discretionary amount of energy when they are connected to it. Moreover, a data plane controls the operation of EPS, where this plane responds to the interaction of loads and energy sources through a data network. A controller using data-plane information decides how much energy is obtained from each of the different energy sources and how much energy is supplied to each load. 

The EPS is capable of integrating the energy supplied by multiple energy sources and deliver this energy to a heavily demanding load. Similarly, the EPS can supply the energy requested by multiple loads from a single energy source. In both scenarios, the aggregation and supply of energy is done in discrete amounts. Because each source-switch or switch-load energy transfer carries a discrete amount of energy, the energy is actually transferred as an energy packet. 

\subsection{Control of Energy Packet Switch}
There are two levels of control for the operation of the switch: a) by a top-level request-grant protocol where loads and energy sources interact with the switch and b) levels of energy in the recipient load and EPS. 

The request-grant protocol is an operation in which all elements of a CDG participate to supply or demand energy \cite{rojas2014testbed}. In short, loads issue a request for the needed energy amounts through a data network (e.g., Internet) and each energy source grants amounts of energy to requesting loads, by issuing notifications through the data network and supply through the grid lines. The amount of granted energy is capped by the feeder capacity. In this framework, loads, energy sources and EPSs are interconnected through the data network, forming an Internet of Things (IoT) environment.

 At the inputs, energy sources send energy packets to the core of the switch. For that, the EPS and the sources execute the request-grant protocol where EPS is the load. At the outputs, EPS supplies controlled amounts of energy as energy packets to the load during a time interval (e.g., a single or multiple time slots). In the latter case, the loads and EPS execute the request-grant protocol, where the EPS is the source. 

To perform these operations, each supercapacitor is connected to an input or an output of switch at any given time slot. EPS has a fully interconnected network where all inputs may be connected to any supercapacitor and each supercapacitor may be connected to any output. In this way, each input (source) may transfer energy to each capacitor. However, only one source can be connected to a single capacitor at a time to avoid undesirable connections between sources.  On the output side, a load may receive energy from one or multiple capacitors. 

Because EPS energy storage is based on supercapacitors, a load also uses a supercapacitor as an interface to receive the granted energy. Energy is then supplied through a  capacitor-capacitor circuit. The advantages of using this approach is that energy transfer is fast and slump sums of energy can be transferred in each opportunity a load is granted to minimize the number of required transfers. The levels of energy transferred are dependent on the size of the used capacitance and the voltage (charge) difference between the source capacitance ($Cs$) at the EPS and the load capacitance ($Cl$) at the load.

For example, the EPS may be modeled as a source capacitor $Cs$ and the amount of charge and energy are given by:
\begin{equation}
Cs=\frac{q_s}{Vs}
\end{equation}
where $q_s$ is the charge held by $Cs$ and $Vs$ is the voltage between its terminals.
The amount of energy $U_s$ in $C_s$ is given by
\begin{equation}
Us=\int_0^{q_s} \frac{q_s}{Cs} d_{q_s}= \frac{1}{2} Cs Vs^2
\end{equation}

It is convenient to have $Cs$ fully charged before any energy transfer to maximize the amount of energy transfer. We consider that $Cs \geq Cl$ to facilitate the flow of energy from $Cs$ to $Cl$. Then, a load may get connected to $Cs$ (i.e., EPS) for energy transfer and during that time, $Cs$ is disconnected from any energy sources. The energy in $Cl$, which is the combined amount of energy before the transfer minus that in $Cs$ after the transfer also depends on how much charge there was in $Cl$ before the transfer. The combined amount of energy in both capacitors after the energy transfer, $U_{sl}$, is:
\begin{equation}
U_{sl}= \frac{1}{2}C_{sl}V_{sl}^2
\end{equation}
where $C_{sl}$ and $V_{sl}$ are the equivalent capacity that includes $Cs$ and $Cl$ and the voltage on the terminals of the capacitors after the energy transfer, respectively. The combined capacitance is modeled as an increased capacitance, $C_{sl}=Cs+Cl$.

The energy difference that EPS may transfer to a load depends on the amount of energy on both capacitors, or:
\begin{equation}
U_{sl}=\frac{1}{2} (Cs+Cl)V_{sl}^2
\label{eq:ul}
\end{equation}
where $V_{sl}=\frac{Cs V_s+Cl Vl}{Cs + Cl}$ and $Vl$ is the voltage on $Cl$ before the energy transfer.

As (\ref{eq:ul}) shows, the amount of energy transferred to $Cl$ depends on the charge in $Cs$ and $Cl$ before the energy transfer occurs. A request carries the value of the charge of $Cl$, and $Cs$ may adapt voltage, capacitance, or a combination of both, according to the amount of energy that is to be granted to $Cl$ in a time slot. In this paper, we use a fixed voltage at $Cs$ before energy is  transferred so that EPS adapts $Cs$ as a discrete value for each amount of transferred energy. In this way, $Cs= n ~Cl$, where $n=\{1, \ldots,~k$\}. In our initial switch prototype, $k=8$.

\section{Examples of Energy Exchange}
\label{sec:testbed}

The operation of the energy packet switch is largely based on the charging process of a capacitor with configurable capacitance. 
Capacitors have the property of charging and discharging at very fast rate, if no large resistance is connected in series to it. Furthermore, the energy density of supercapacitors has been recently increased such that the amount of energy that can be stored in today's supercapacitors is becoming applicable to higher-power loads. 

The proposed switch uses supercapacitors to collect energy from one or multiple sources. This source could be an alternative energy source or batteries charged by it or another EPS. Here, we show an example of the charging operation of a supercapacitor.

\begin{figure}[htb]
\centering
\includegraphics[width=2.5in]{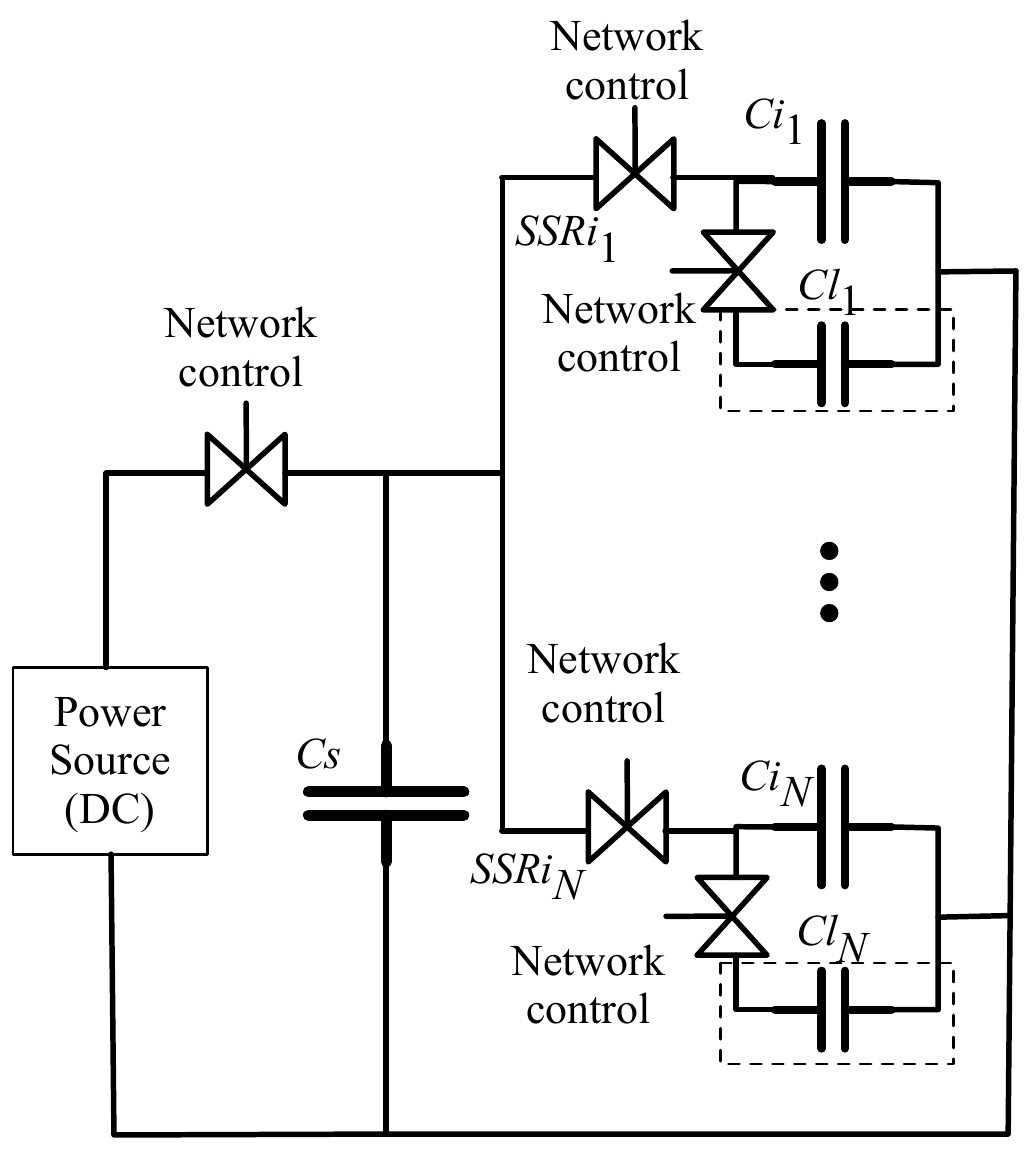}
\caption{Testbed to transfer discrete energy to capacitive loads (half cycle).}
\label{fig:transfer-to-N-loads}
\end{figure}

Figure \ref{fig:transfer-to-N-loads} shows a simplified circuit of the EPS for transfer energy from the switch  (i.e., $Cs$) to $N$ separate loads. Here, SSRs are used as controllable switches that enable charging a supercapacitors or passing energy from the supercapacitors in the EPS to the superpacitor interface ($Ci$) that is connected to requesting loads (using $Cl$ as receiving capacitor). The gate of an SSR is network controlled; meaning that a PAP enables the SSR (labeled as ``Network control'' in the figure) after information is received from the data network. 
In this example of energy transfer, we show the connection when $Cs=Cl$ (i.e., a single source capacitor to a single load capacitor). 

For a fast-paced energy transfer, current limiters based on passive resistance are avoided and an energy source with high-current capacity may act as a fast-charging supply to rapidly charge the capacitors in the EPS. In turn, the EPS may act as a fast-charging-discharging device. Although loads may also receive charge at a fast rate (via $Cl$), they may consume energy at slower rates. Although we aim for an fast-charging-discharge EPS, controlling the energy transfer speed is left out of the scope of this paper.

\begin{figure}[htb]
\centering
\includegraphics[width=2.5in]{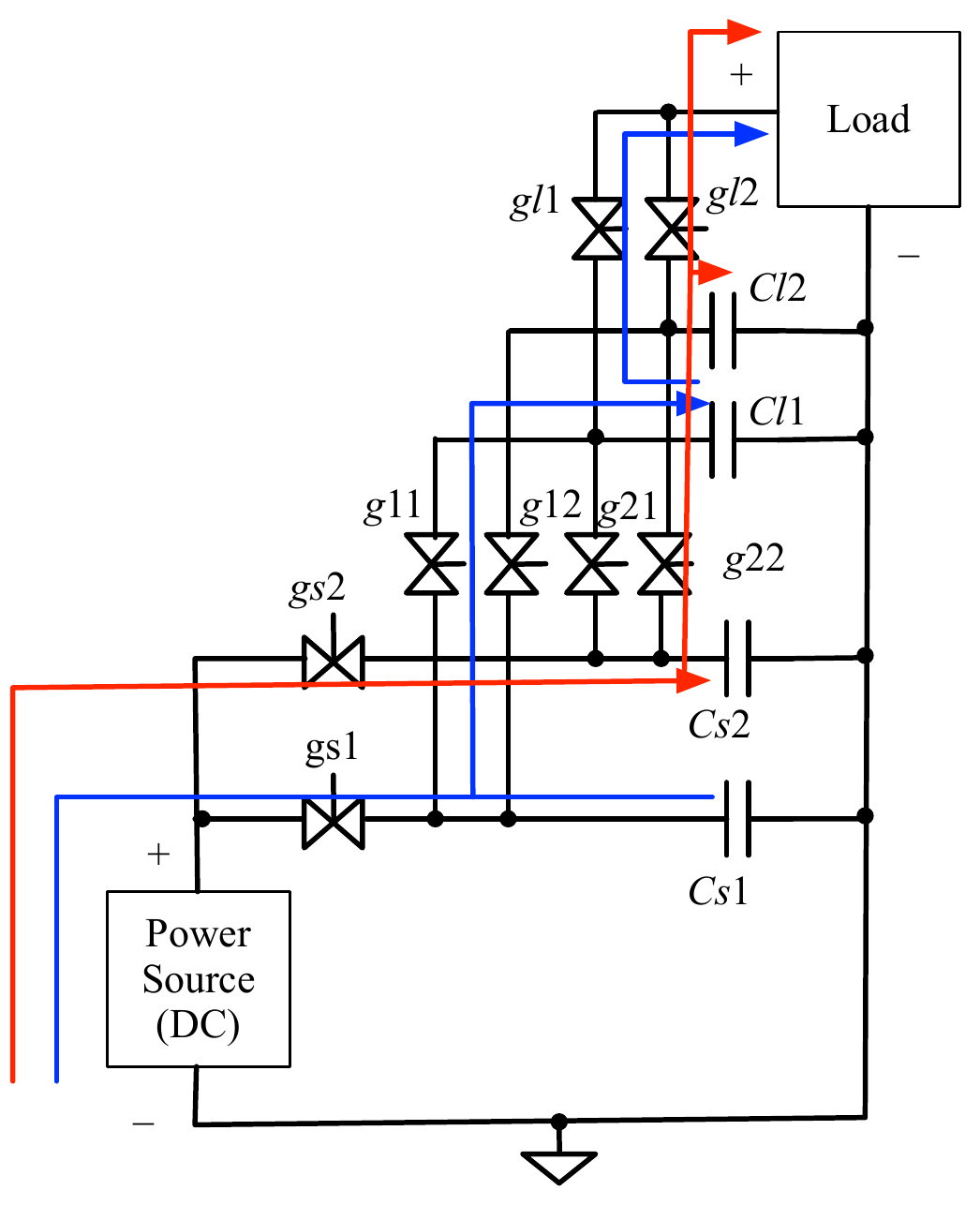}
\caption{Testbed of continuos transfer of discrete energy to a load (full cycle).}
\label{fig:transfer-continous}
\end{figure}

The last figure showed a circuit that charges one or multiple loads but in such an approach the load capacitor can either receive or supply energy (half duplex interface). We can overcome that limitation by increasing the number of components, as Figure \ref{fig:transfer-continous} shows. In this diagram, there are two $Cl$s, where each of them exchange roles (charging/discharging) for a continuous energy supply. In this circuit, there are two interface capacitors for the load. Each capacitor may charge during alternating half cycles and discharge in the same fashion. This is, while one capacitor charges, the other supplies energy to the load, and vice versa. To achieve this, the circuit has twice the number of load components as compared to that in Figure \ref{fig:transfer-to-N-loads}.
To simplify the description of the figure, the drawing includes two paths (blue and red lines) followed by the energy flow from $Cs$ to one of the interface capacitors ($Ci$) and towards $Cl$.  A similar operation occurs with the $Cs$ and $Ci$s. As the figure shows, $Ci$ supercapacitors are used as interface between $Cs$ and $Cl$ in this approach. In this way, $Cs$ first transfers energy to $Ci$s and then each $Ci$ transfers energy to the corresponding $Cl$.

\section{Experimental Results}
\label{sec:results}

\begin{figure}[!t]
\centering
\includegraphics[width=3.5in]{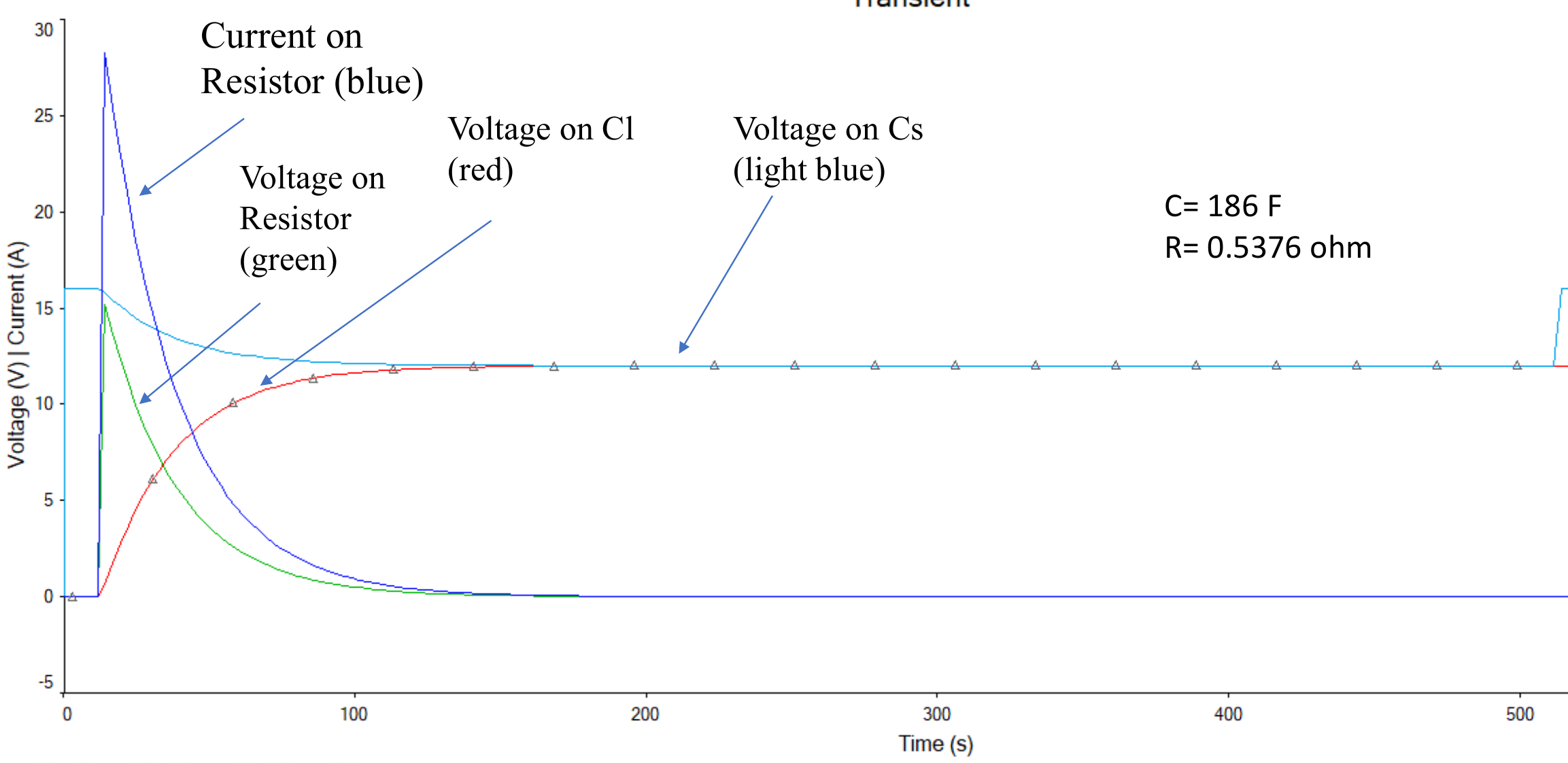}
\caption{Current and voltage response on energy transfer to load capacitor $Cl$ in a resistive circuit.}
\label{fig:CR-transient}
\end{figure}

In this section, we present simulation experiment of the energy exchange process. Simulation of the presented circuits is performed with NI Multisim $\copyright$. First, we show that energy may be exchanged between two capacitors ($Cs$ and $Cl$) to perform a controlled energy delivery.  Figure \ref{fig:CR-transient} shows the charging process of a capacitor when it is exposed to an unlimited amount of energy (i.e., current) but with a voltage set to 12 V. As the figure shows, the capacitor gets a charge proportional to the voltage and charges at speed paced by the $RC$ constant, where $R$ is the resistance on the circuit and $C$ the capacitance of the (charge receiving) capacitor ($Cl$).

\begin{figure}[htb]
\centering
\includegraphics[width=3.5in]{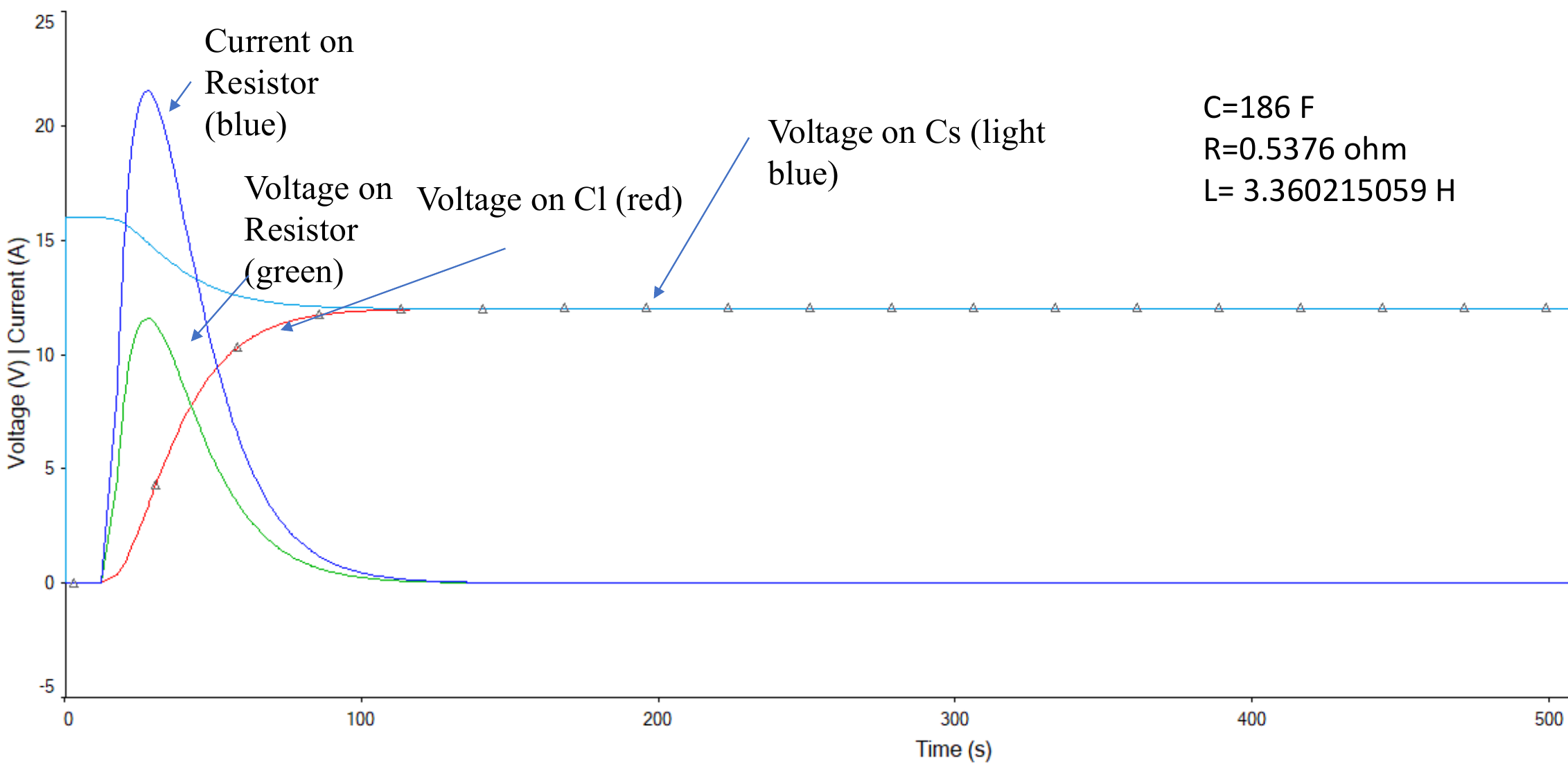}
\caption{Current and voltage response on energy transfer to load capacitor $Cl$ in a resistive circuit.}
\label{fig:CLR-transient}
\end{figure}

As the figure shows, an ideal capacitor is charged at a very fast rate under an energy source with unlimited current capacity but the amount of current that circulates through the circuit is also very large and so is the amount of dissipated power. This intense-power transfer may heat the interconnection as time passes. Therefore, we are interested in managing the amount of current that can be transferred in the energy packet switch and for that we may add an inductor, connected in series with the load (and capacitor). This circuit may reduce energy loses that otherwise may occur with the use of a resistive limiter. Figure \ref{fig:CLR-transient} shows a charging process in such a Resistance-Capacitance-Inductance (RCL) circuit. This figure also shows that the amount of current used to charge the capacitor effectively decreases but the presence of the inductance generates oscillations that need to be considered in the estimating the charge of the capacitor. 

\begin{figure}[htbp]
   \centering
    \subfigure[Energy transfer from $Cs$ to $Cl$ for different initial voltage at $Cl$.]{
   \includegraphics[width=3.4in]{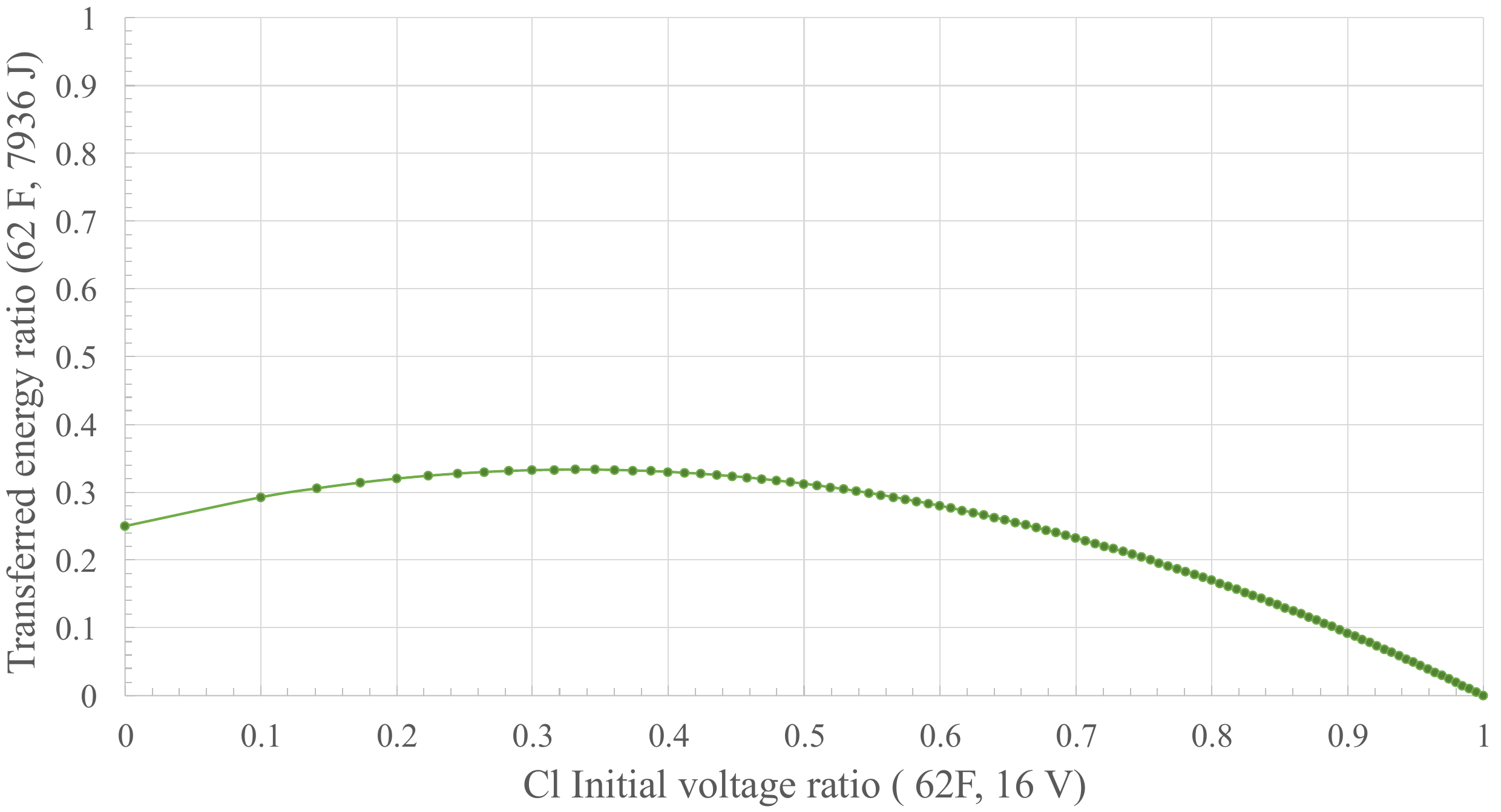}
   \label{fig:exchange-voltage-cs-cl}}
\subfigure[Energy transfer from $Cs$ to $Cl$ for different initial energy at $Cl$.]{
   \includegraphics[width=3.4in]{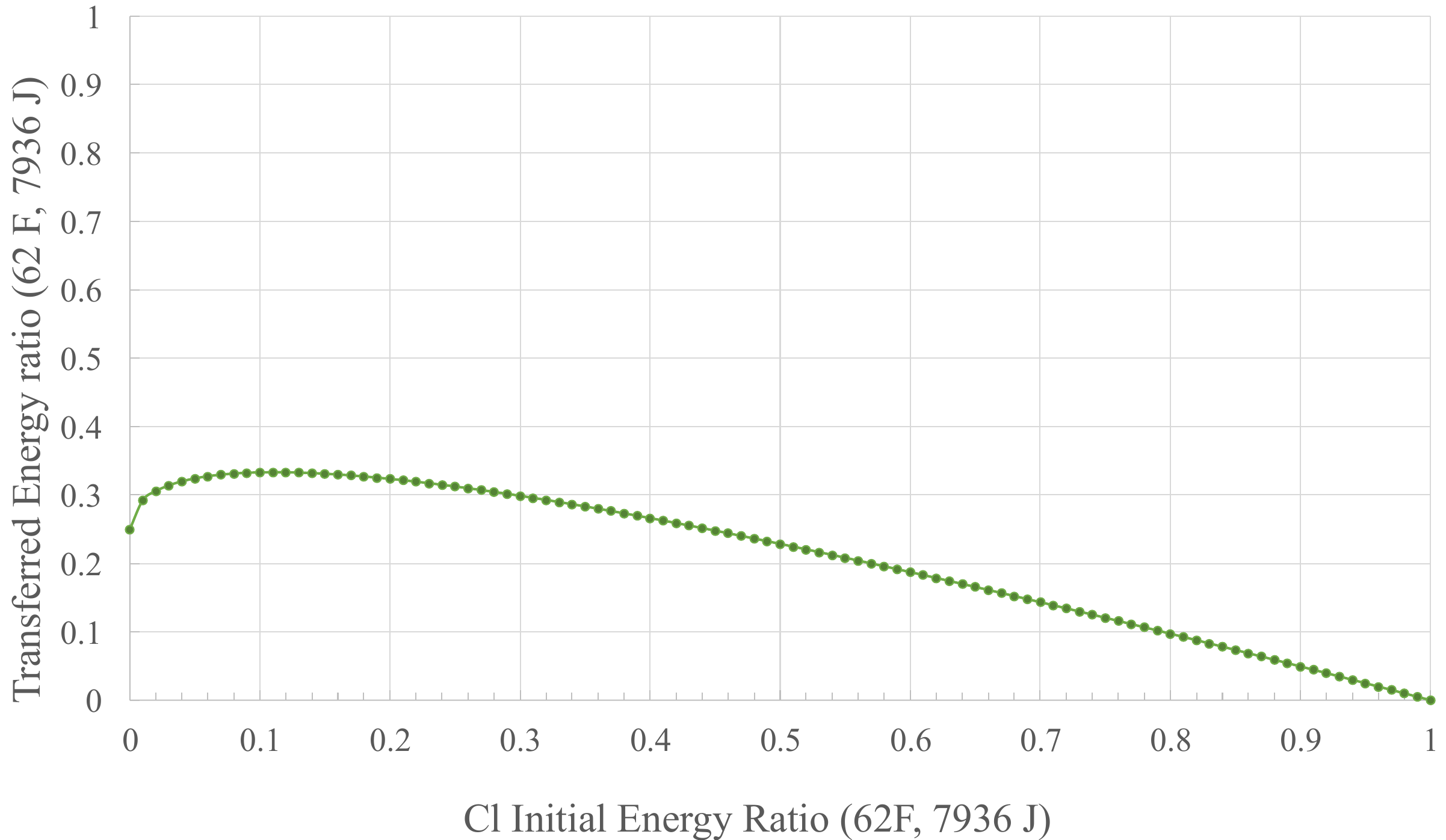}
   \label{fig:exchange-cs-cl}}
   \caption{Energy transfer for a $Cs=Cl$ for $Cs=12$ V as initial voltage.}
 \label{fig:transfer-cs-cl}
\end{figure}

For the exchange of energy between $Cs$ and $Cl$, we consider different size ratios in a capacitive circuit. Specifically, the $Cs-Cl$ ratio is $\frac{Cs}{Cl}$ or number of times $Cs$ is larger than $Cl$.  Figure \ref{fig:exchange-voltage-cs-cl} shows the voltage changes on $Cl$ for a $Cs$ always fully charged (voltage of $Cs$ equals that of the source) as initial condition.

Figure \ref{fig:exchange-cs-cl} shows the amount of energy transferred from $Cs$ to $Cl$ when their capacitance ratio is 1 (i.e., $Cs=Cl$) and, with $Cs$ charged to 12 V and for $Cl$ holding different level of charge (measured in volts through its terminals). As the graph shows, the maximum transfer of the energy is achieved when $Cl$ has a small charge. Therefore, showing an optimal point. As expected, as the charge of $Cl$ increases, the amount of transferred charge decreases. This slowdown on the amount of energy charge is due to  decreasing difference of voltages between $Cs$ and $Cl$.

\begin{figure}[htbp]
   \centering
    \subfigure[Energy transfer from $Cs$ to $Cl$ for different initial voltage at $Cl$.]{
   \includegraphics[width=3.4in]{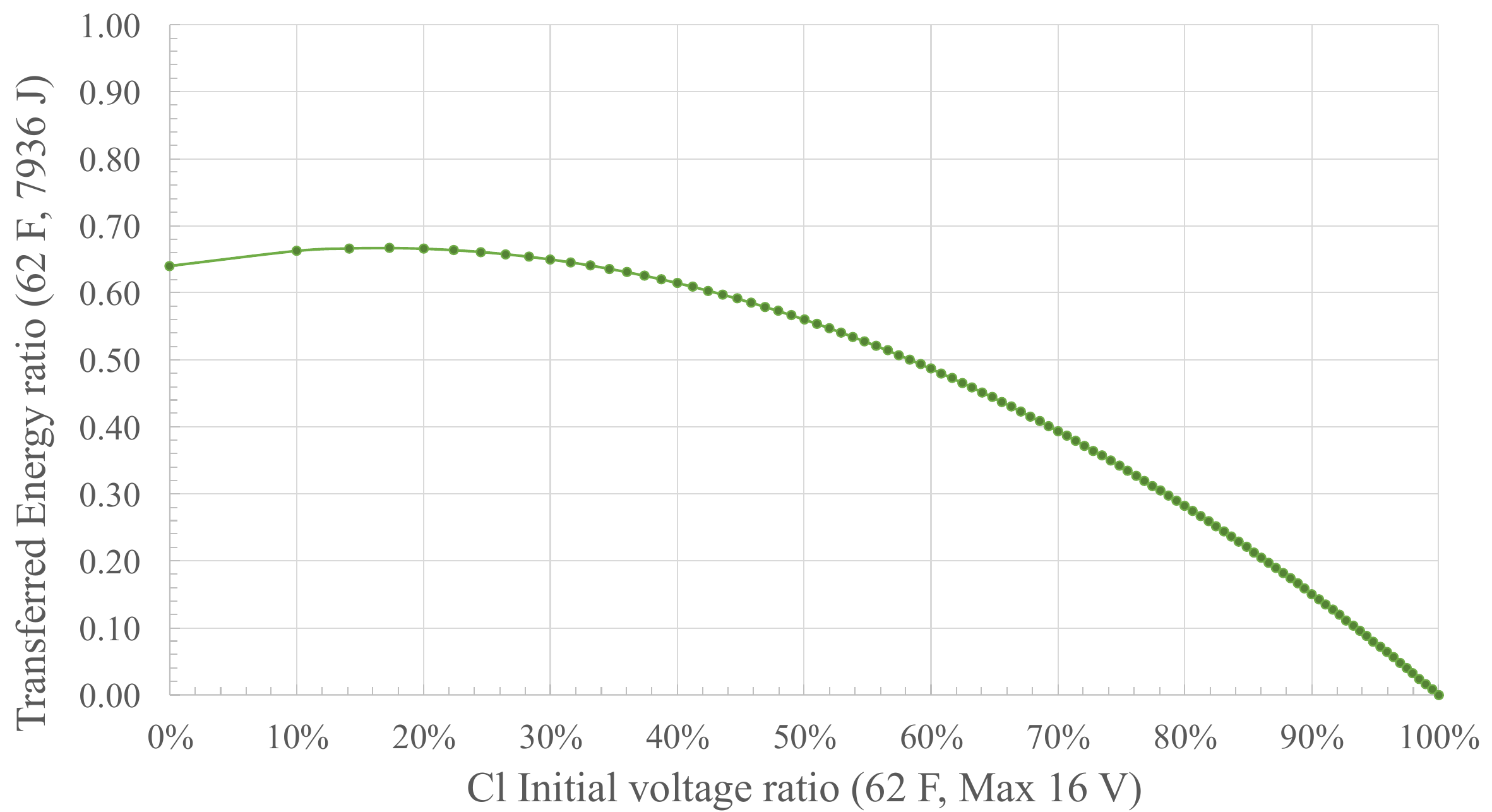}
   \label{ffig:exchange-voltage-cs-4cl}}
\subfigure[Energy transfer from $Cs$ to $Cl$ for different initial energy amount at $Cl$.]{
   \includegraphics[width=3.4in]{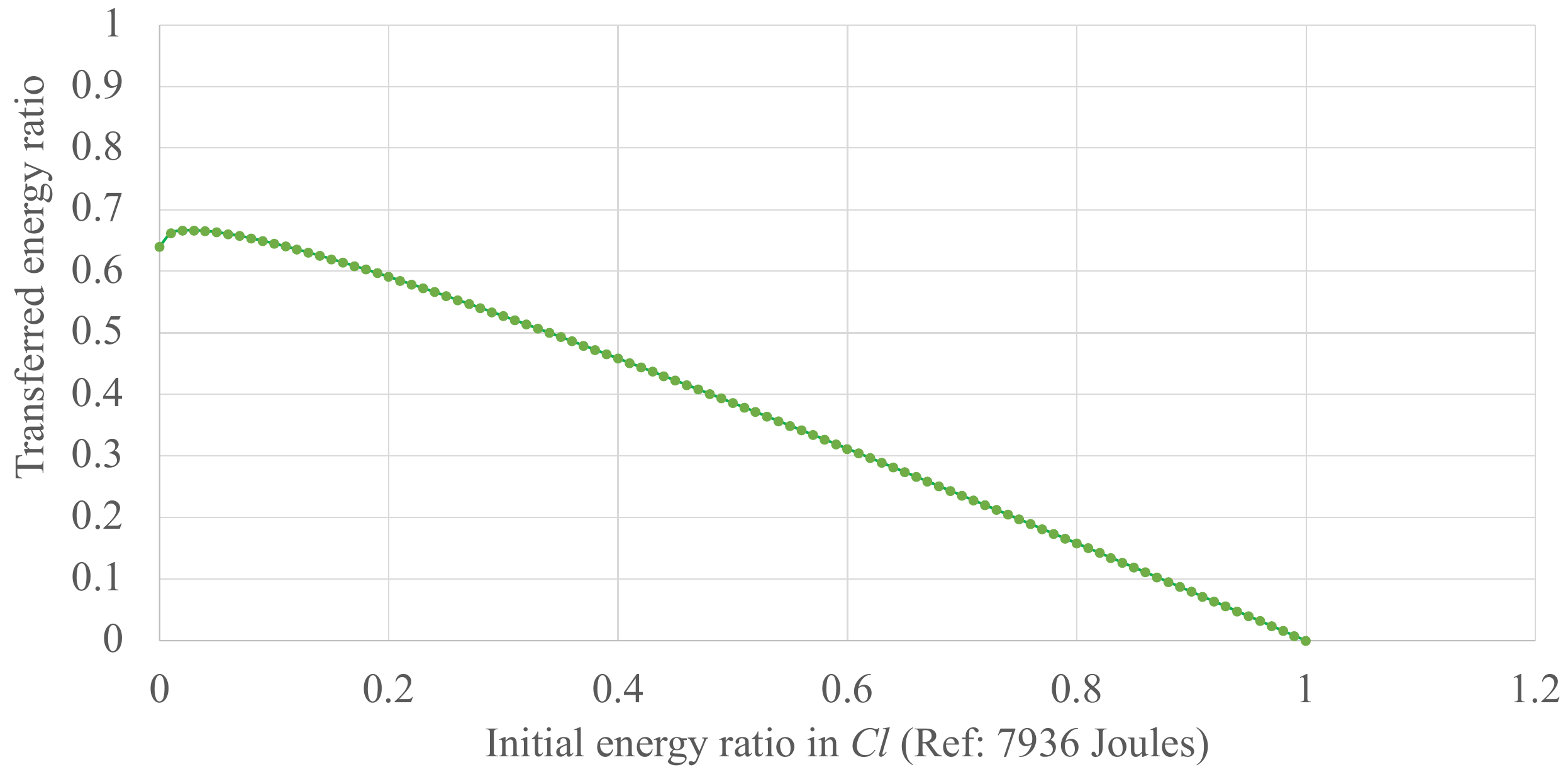}
   \label{fig:exchange-cs-4cl}}
   \caption{Energy transfer from $Cs$ to $Cl$ for $Cs=4Cl$.}
 \label{fig:transfer-cs-cl}
\end{figure}

Figure \ref{fig:exchange-cs-4cl} shows similar graph as before but for a capacitance ratio $\frac{Cs}{Cl}=4$. In this case, the trend of the transferred energy is similar to the previous case, but the largest amount of transferred energy ratio is now about 0.67  This ratio is about twice the energy transferred as compared to that for $Cs=Cl$. This is, a larger $Cs$ capacitance allows for a larger transfer of charge to the load capacitor.

\begin{figure}[htb]
\centering
\includegraphics[width=3.5in]{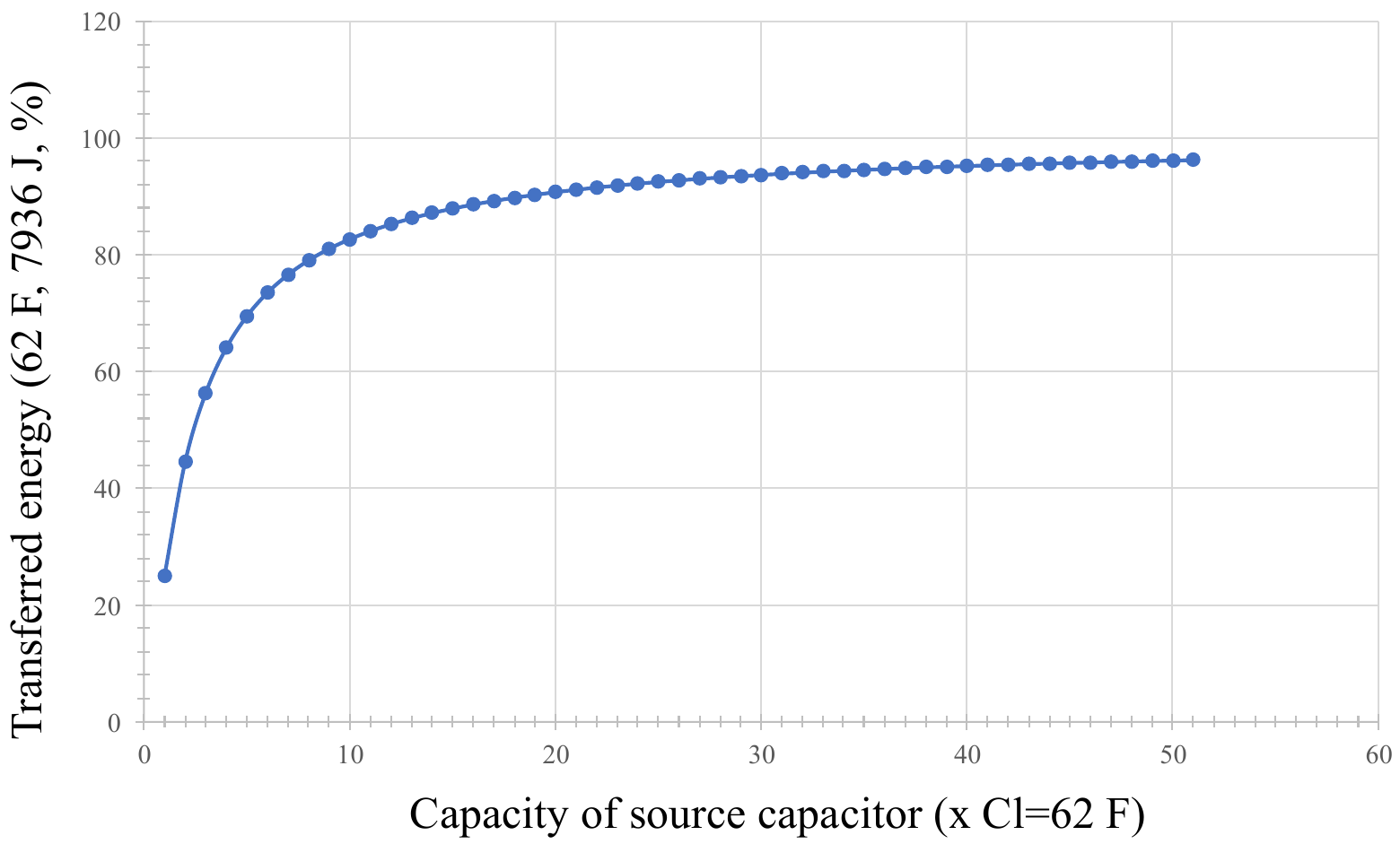}
\caption{Energy transfer from Cs to Cl as a function of Cs capacity.}
\label{fig:energy-transfer-vs-Cs-capacity}
\end{figure}

Figure \ref{fig:energy-transfer-vs-Cs-capacity} shows a general case of energy transfer ratio for general $Cs$ capacitance vs $Cl$, but for a discharged $Cl$ as initial condition. As expected, the amount of energy transfer increases as the capacitance of $Cs$ increases. However, the mayor contributions of transferred energy occur for $Cs=4Cl$. Although the increase of $Cs$ also increases the amount of transferred energy, this increase becomes less significant for larger ratios. Note that for charging $Cl$ to close to 100\% (approaching to $Vs$), it would require a very large $Cs$ capacitance. 

\begin{figure}[htb]
\centering
\includegraphics[width=3.5in]{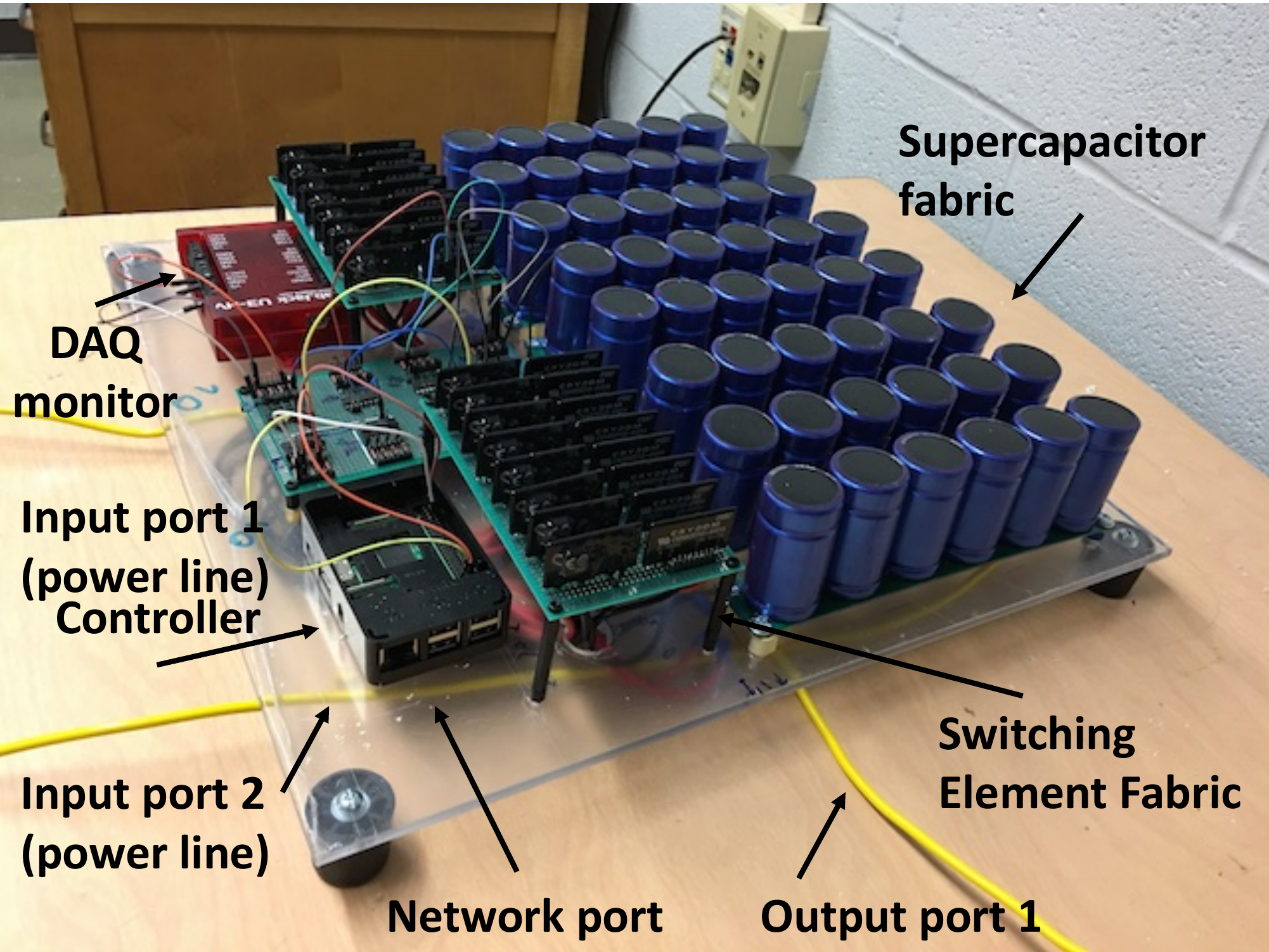}
\caption{Photograph of our EPS prototype.}
\label{fig:EPS-photo}
\end{figure}

Figure \ref{fig:EPS-photo} shows a photograph of our EPS prototype. The switch has eight supercapacitor arrays, where each capacitor array is a set of five supercapacitors to enable using up to 16 V. There is a switch controller, an array of switching elements, and a digital acquisition board (DAQ) used to monitor the state of the supercapacitors. The figure shows two power inputs and one (of the two) power outputs. In addition, the switch has a network interface to communicate with compatible sources and loads.

\section{Conclusions}
\label{sec:conclusions}
We proposed what we believe is the first energy packet switch able to cope with the demands on energy transfers of a digital power grid. The design of this energy packet switch is based on shared capacitance; supercapacitors may be shared by either energy sources or by loads at any given time.
The operation of the proposed switch is similar to that of a packet switch used for the transmission of data in the Internet. Energy is received and aggregated through energy storage and energy storage capacity is discretely (digitally) adjusted. The buffers in the energy packet switch are implemented with supercapacitors in this switch but other forms of energy storage, batteries, inductors, etc., can also be considered. In this example, supercapacitors can receive and transfer energy very quickly, with high efficiency, and are able to provide energy in bound amounts that give place to energy quanta or energy packets. We show examples of the transfer of energy between the EPS and load capacitances. In addition, we show a capacitor circuit combined with an inductance to curve transfers by intense currents. 
The energy packet switch has a control plane where data transmitted is used to control the operation of the switch, and a power plane to receive and transmit energy between ports. The power and data planes work in parallel.  The data plane follows a request-grant protocol, where loads issue request packets through the data network and energy transfers are granted and notified through the same network.  
The energy packet switch also provides the important ability of integrating different energy sources. Therefore, the EPS also provides a method to integrate alternative energy sources without overwhelming and unbalancing the grid.\\

\noindent{\bf Disclaimer:} Any opinions, findings, and conclusions or recommendations expressed in this material are those of the author(s) and do not necessarily reflect the views of the National Science Foundation.

\end{document}